# Performance Analysis and Fault Diagnosis Method for Concentrator Photovoltaic Modules

Harsh G. Kamath, Nicholas J. Ekins-Daukes, Kenji Araki, and Sheela K. Ramasesha

*Abstract*— Concentrator Photovoltaic (CPV) systems use high efficiency multi-junction solar cells with efficiencies >40%, but the module efficiency is often much lower. The increased complexity of a CPV module, with optics, receiver and the tracker give an increased probability that faults will arise during the operational lifetime. In addition, a location like India has varied atmospheric conditions that further complicate the diagnosis of faults. It is therefore important to decouple effects due to the external environment (such as the atmosphere) from effects due to the degradation of the module. By applying a computer model to outdoor CPV test data in Bangalore, India we have established a method to assess the performance of the CPV module and finally we present a method to diagnose faults in the module.

## I. INTRODUCTION

CONCENTRATOR photovoltaic systems offer the possibility for low cost electricity generation in regions with high values of Direct Normal Irradiance (DNI). However, the use of spectrally sensitive multi-junction (MJ) solar cells complicate the process of assessing the performance of the module since the system will be dependent upon the spectral distribution of sunlight, not just the total irradiance. India, where the modules are installed, has particularly varied atmospheric conditions, both in terms of aerosol distribution and precipitable water [1]. It is therefore important to determine the effects of these factors on the electricity yield and ensure that an investment in CPV technology in India will return the expected quantity of electricity. A pilot study was therefore performed at the Divecha Centre for Climate Change, Indian Institute of Science, Bangalore, where annual DNI is around 5–5.5 KWh/m$^2$/day.

The performance of CPV module is influenced principally by the tracking accuracy, cell temperature and the incident spectrum. Any difference in the spectrum is first reflected in the short-circuit current ($I_{sc}$) followed by the fill-factor and evaluated extensively by Meusel et al. [2]. Araki et al., who considered the case for spectral mismatches in Nagoya, Japan [3] while Hashimoto et.al made a comparative study between similar systems located in Aurora, USA and Okayama, Japan. They found that the spectrum is the key factor that is responsible for reduction in performance during the winter season at both the locations [4]. More recently, Chan et al. carried out a detailed study on effects of individual atmospheric parameters on energy yield [1, 5]. The effects of variation in spectral irradiance on MJ CPV systems is reported in [6].

Previously, regression models have been used, empirically correlating various atmospheric parameters along with the power output. A review of various previously used methods for yield estimation can be found in [1]. Alternatively, a bottom-up model that simulates the performance of the MJ cell, module optics and system using an equivalent circuit [7] has been validated for a module located in Toyohashi, Japan to estimate the annual yield which had an excellent agreement with experimental data [8].

Here we extend a 'bottom-up' equivalent circuit model to analyse a CPV module located in Bangalore, India. Section II in this paper characterises the module behaviour and the following sections are on modelling and analysis of solar spectrum and CPV module, leading to fault diagnosis method in section V. The analysis of the module is complicated by the presence of a wide distribution of short circuit current of cells in the module which we have identified and characterised by comparing simulated and measured data under different atmospheric conditions. The present paper extends this analysis to the problem of identifying additional series resistance that have arisen during operation. In flat plate PV, degradation may be determined through IV parameters [9, 10] and visual inspection [11] and is found to depend on the geographical locations where the modules are installed [12]. Some methodologies exist to detect the faults like string disconnections in flat plate PV systems from a comparison of the ac power output recorded and estimated ac power from a model [13], however it is generally more informative to use IV

This work was partially supported by the European commission and NEDO through the funding NGCPV (EU Ref.N: 283798). (*Corresponding author: Sheela K. Ramasesha*)

H. G. Kamath is with Divecha Centre for Climate Change, Indian Institute of Science, C. V. Raman Ave, Bangalore-560012, India (e-mail: harsh.kamath@hotmail.com, harshkamath@iisc.ac.in).

N. J. Ekins-Daukes, was with Department of Physics, Imperial College, London, SW7 2AZ, UK. He is now with School of Photovoltaics and renewable Energy Engineering, University of New South Wales, Kensington, Australia, NSW 2052 (e-mail: nekins@unsw.edu.au).

Kenji Araki is with Toyato Technological Institute, 2-1-2 Hisakata, Tempaku-Ku, Nagoya, Japan (e-mail: cpvkenjiaraki@yahoo.co.jp).

S. K. Ramasesha was with Divecha Centre for Climate Change, Indian Institute of Science, C. V. Raman Ave, Bangalore-560012. She is now with National Institute of Advanced Studies, Indian Institute of Science campus, Bangalore-560012, India (e-mail: sheela.ramasesha@gmail.com).

curve data to both identify module degradation [14] and the nature of the degradation. By fitting the IV curves under different irradiance conditions, a model for the faulty module was constructed and provides insight into the mode of degradation. It is necessary to measure the IV curves and have meteorological data (refer section 3) to use this method.

## II. MODULE DESCRIPTION AND LOCAL ENVIRONMENT

A 820X CPV module with 25 triple junction InGaP/GaAs/Ge solar cells connected in series was supplied by Daido Steel and is mounted on a dual axis tracker by Green Source Technology. The acceptance angle of the CPV receiver is ±0.92 degree and recommended tracking accuracy is ±0.7 degree while the tracker accuracy is less than ±0.2 degree. A cell efficiency of 38.5% was achieved at Standard Test Condition (STC) and module efficiencies of 28% and 26% with fill-factors 83.4% and 82.8% were achieved at STC and Standard Operating Condition (SOC), respectively [15,16]. The CPV receiver has a glass homogenizer that reduce the possibility of chromatic aberration. The band gap combination of the 3-junction cell at 298 Kelvin is 1.89 eV, 1.42 eV and 0.66 eV [8]. A schematic of the CPV receiver is shown in figure 1.

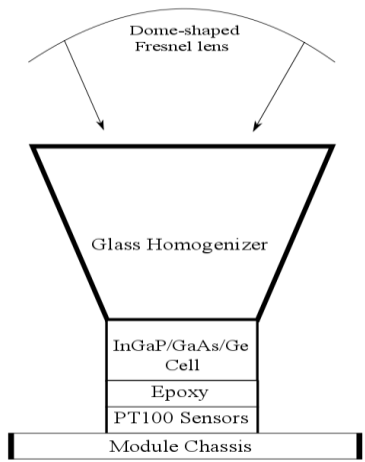

Fig. 1. Schematic of the CPV receiver

DNI was measured using a Kipp & Zonen CHP1 pyroheliometer and cell temperatures were sensed using PT100 temperature sensors embedded in the module under each solar cell. The temperature sensors do not influence the performance of the cells [17]. The IV curves were measured every ~20 seconds. A weather monitoring system Vintage Pro 2 supplied by Davis Instruments that measures the ambient temperature, wind speed, wind direction, relative humidity and atmospheric pressure was also used.

Figure 2 shows the DNI profiles for 7$^{th}$ January 2014, 20$^{th}$ April 2014 and 18$^{th}$ November 2014. 7$^{th}$ January 2014 was a clear day with no cloudiness and a peak DNI of around 950 W/m$^2$ is recorded at around 11AM. January is winter in Bangalore and is a time of year when India has little cloud cover and receives high DNI; in contrast to Spain and South Africa where high DNIs are received in summers [18]. April is pre-monsoon and November is post monsoon. During the months of April and November, we see intermittent cloudiness, especially during afternoon and evening.

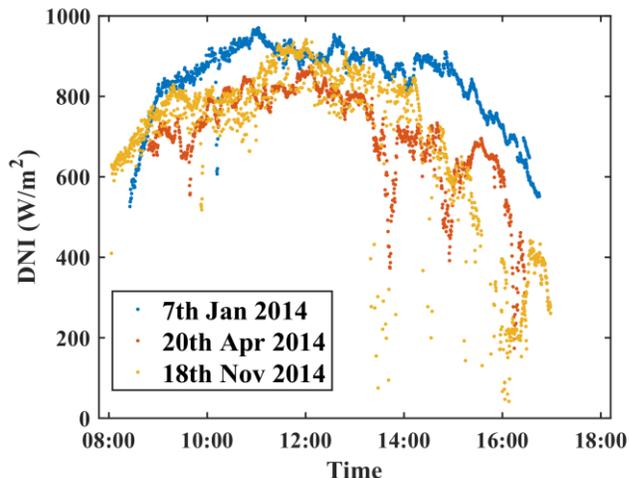

Fig. 2. DNI profile on 7$^{th}$ January 2014, 20$^{th}$ April 2014 and 18$^{th}$ November 2014

The thermal behaviour of the module was verified using PT100 sensors located in a matrix across the module. The module is found to heat up quickly but cool down slowly. Figure 3a shows the temperature of channels 7, 10 and 12 in the module on 7$^{th}$ January 2014. The temperatures and corresponding values of open circuit voltage ($V_{oc}$) of individual cells at 11 AM on 7$^{th}$ January is shown in figure 3b. The individual cell temperatures within the module are important as the cell temperature is inversely proportional to $V_{oc}$ and the module $V_{oc}$ is the sum of individual cell $V_{oc}$ values. The value of $V_{oc}$ for individual cells are obtained using a computer model that has been optimised for the cells in our module (discussed later in section 4). We learn that the cell temperatures are higher at the module centre and lower at the module edges. This is consistent with the study conducted by Ota et.al [19] and Castro et.al [20]. The cell temperatures were further correlated with the wind speed and it is concluded that the temperature variation is small for the wind speed at the test location and wind speed did not have a huge impact on performance.

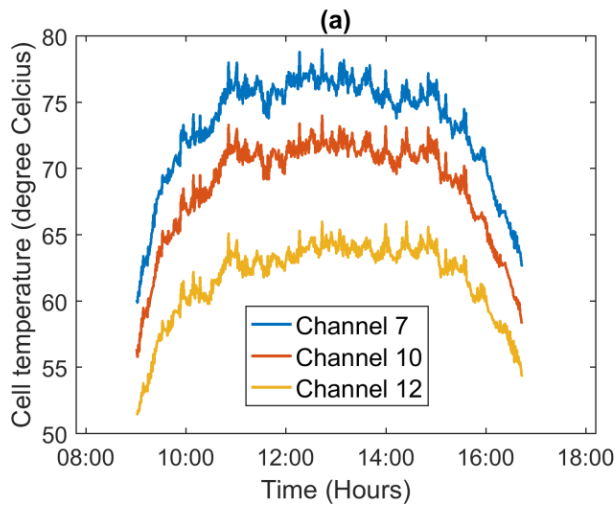

| Channel: 1 | | Channel: 2 | | Channel: 3 |
| 69.6 °C | | 74.8 °C | | 73.3 °C |
| 2.632 V | | 2.589 V | | 2.602 V |
| | Channel: 4 | | Channel: 5 | |
| | 77.4 °C | | 79 °C | |
| | 2.568 V | | 2.555 V | |
| Channel: 6 | | Channel: 7 | | Channel: 8 |
| 69.2 °C | | 78 °C | | 73.5 °C |
| 2.635 V | | 2.563 V | | 2.6 V |
| | Channel: 9 | | Channel: 10 | |
| | 70 °C | | 73 °C | |
| | 2.629 V | | 2.604 V | |
| Channel: 11 | | Channel: 13 | | Channel: 12 |
| 72.3 °C | | 74.4 °C | | 64.8 °C |
| 2.61 V | | 2.5933 V | | 2.671 V |

Fig. 3. (a) Cell temperatures within module on 7th January 2014 and (b) instantaneous cell temperatures and open-circuit voltages at 11 AM, 7th January 2014

## III. THE SPECTRUM

To assess the performance of CPV module using equivalent circuit model simulations, it is necessary to produce the incident spectrum correctly as the sub-cell photocurrents heavily depend on it and also to decouple the effect of spectral difference from actual CPV performance. While the spectrum is seldom measured, the meteorological parameters such as Aerosol Optical Depth (AOD) and precipitable water (PW) are available (from AERONET or other ground-based measurements). Hence, we can simulate the CPV system by modelling the spectrum using AOD and PW as prominent parameters with a radiative transfer model, as described in this work. Importantly, AOD and PW were not available for the site for spectrum modelling and the measured spectrum had spectral irradiance measurements only up to 1050 nm. Hence a methodology was developed using two irradiance models (SMARTS2 [21] & SPCTRAL2 [22]) to estimate these from the measured spectrum and DNI data. The PW was estimated by fitting the measured and calculated spectrum in the regions where water absorption leads to dips in the solar spectrum, notably at 930 nm with SPCTRAL2 model and is used as an input to SMARTS2 to model the entire spectrum. The Shettle and Fenn (S&F) urban aerosol model in SMARTS2, which considers the effect of relative humidity on aerosols, was chosen to best match the short wavelength response and the AOD is adjusted so that the integrated spectral irradiance matched the measured DNI.

A few clear days were picked from the month of January 2014 and the spectrum was fitted using SPCTRAL2 and SMARTS2. The value of PW and AOD were noted from the fitted spectrum. Table 1 presents the parameters used for fitting the spectrum and table 2 shows the calculated and measured parameters from fitted spectrum for 7th of January 2014 which was a clear day and variation in AOD over the day was not much. The prominent parameter that effects the performance of a CPV system during such a day is airmass. Figure 4 illustrates the effects of AOD and PW on the spectrum. We can clearly see that the shorter wavelengths of the spectrum are mainly determined by the AOD, while PW leads to absorption features in the near-infrared.

TABLE I
PRINCIPLE PARAMETERS USED FOR THE CALCULATION OF SPECTRAL IRRADIANCE

| Parameter | Model | Remark |
|---|---|---|
| Aerosol Model | S&F Urban Aerosol Model in SMARTS2 | Angstrom exponent calculated based on relative humidity |
| Ambient temperature | SMARTS2 | From weather monitoring system |
| Relative humidity | SMARTS2 | From weather monitoring system |
| Atmospheric pressure | SMARTS2 | From weather monitoring system |
| Aerosol Optical Depth (AOD) | Calculated with SMARTS2 at 500 nm | Fitted for shorter wavelength response |
| Precipitable Water (PW) | Calculated with SPCTRAL2 | Fitted for near infrared dips, notable 930 at nm |
| Spectral range | SMARTS2 | 280 nm to 2500 nm |

TABLE II
VARIOUS COMPUTED AND MEASURED PARAMETERS FROM SPECTRUM FITTING ON JANUARY 7, 2014

| Time | Relative humidity (%) | AOD at 500 nm | PW (cm) | Measured DNI (W/m$^2$) | Modelled DNI (W/m$^2$) |
|---|---|---|---|---|---|
| 9:00 | 81 | 0.08 | 2.7 | 825 | 820 |
| 10:00 | 67 | 0.1 | 2.57 | 880 | 882 |
| 11:00 | 45 | 0.075 | 2.01 | 960 | 964 |
| 12:00 | 43 | 0.176 | 2.11 | 912 | 914 |
| 13:00 | 35 | 0.164 | 1.906 | 898 | 895 |
| 14:00 | 33 | 0.1811 | 1.88 | 856 | 864 |
| 15:00 | 30 | 0.125 | 1.74 | 858 | 857 |
| 16:00 | 33 | 0.153 | 1.9 | 726 | 727 |

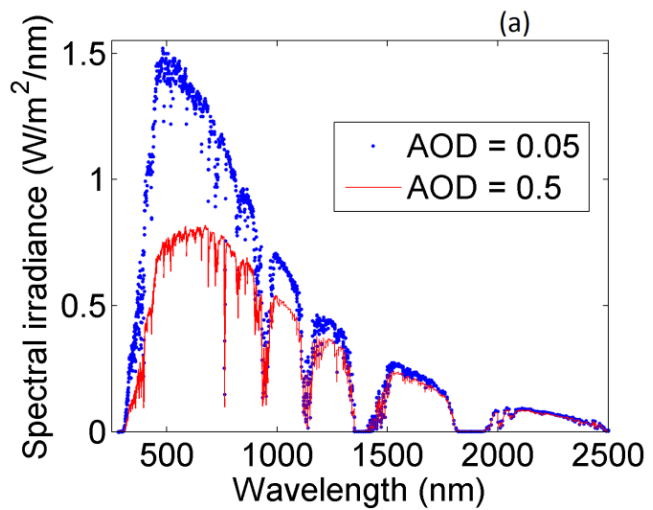

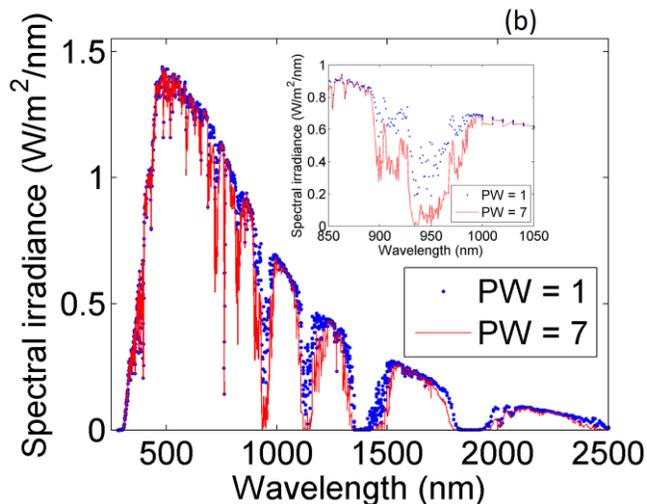

Fig. 4. Effects of (a) AOD at 500nm with airmass = 1.5 and PW = 1.42 cm (b) PW with airmass = 1.5 and AOD at 500 nm = 0.084 on spectrum. The values 0.084, 1.42 and 1.5 for AOD at 500nm, PW and airmass respectively are used as they are standard values for AM 1.5D spectrum. The spectrums are simulated with SMARTS2.

The spectral irradiance measured on the 7th of January 2014 at 1 PM is shown in figure 5 along with the SMARTS2 modelled spectrum. The radiative transfer models work during clear skies. To account for cloudy conditions, a simple approximation could be made by linearly scaling the spectrum such that the measured DNI equals the integrated spectrum, with spectral shape maintained.

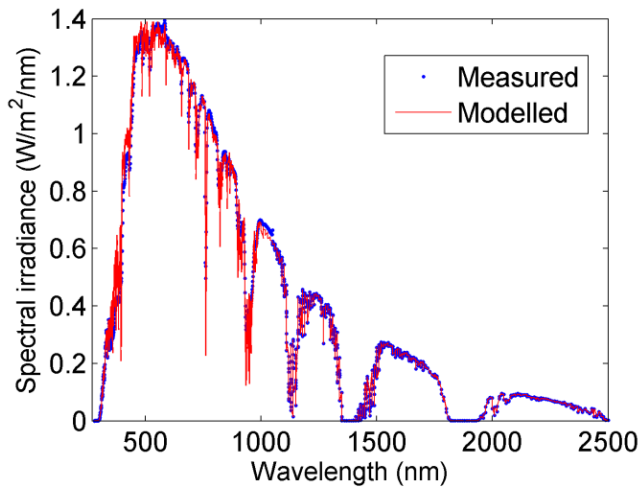

Fig. 5. Measured and modelled spectrum on 7th January 2014 at 1 PM

The atmosphere above Bangalore is complex and some uncertainty arises in matching the measured data to the spectrum. The difference in short circuit current with the SMARTS2 modelled spectrum and the measured spectrum is shown in table 3, showing an absolute uncertainty of about 5 percent, even when the DNIs are in close agreement. This error is likely to arise through a combination of spectral irradiance and DNI measurement uncertainties.

TABLE III
COMPARISON OF SMARTS2 CALCULATED SHORT-CIRCUIT CURRENT AND DNI WITH MEASURED VALUES ON 7TH JANUARY 2014

| Time | Calculated DNI (W/m²) | Measured DNI (W/m²) | % change in DNI | $I_{SC}$ with SMARTS2 spectrum (A) | $I_{SC}$ with measured spectrum (A) | % change in $I_{SC}$ |
|---|---|---|---|---|---|---|
| 11:00 | 963.9 | 960.2 | 0.39 | 3.54 | 3.37 | 5 |
| 12:00 | 914 | 912.5 | 0.16 | 3.47 | 3.32 | 4.5 |
| 13:00 | 895.5 | 897.8 | -0.25 | 3.32 | 3.24 | 2.5 |
| 14:00 | 863.9 | 855.7 | 0.95 | 3.16 | 3.12 | 1.3 |
| 15:00 | 856.6 | 858.3 | -0.20 | 3.02 | 3.08 | -1.9 |
| 16:00 | 727.2 | 726.1 | 0.15 | 2.51 | 2.43 | 3.3 |

## IV. DESCRIPTION OF SOLCORE

A computational framework for evaluating the performance of multi-junction solar cells has been developed called SolCore [23]. Written in the Python3 programming language it enables multi-junction solar cells to be simulated from the bottom up, starting with fundamental semiconductor properties of the component materials through to the full device which is ultimately encapsulated by a double-diode equivalent circuit which in turn enables arrays of cells to be configured into a module [8] [24]. The process is shown in figure 6(a).

SolCore interfaces to the SMARTS2 radiative transfer code enabling the variations in atmospheric parameters discussed earlier to be accounted for. To model the individual cell, cell temperature at the test site and cell parameters like cell area, layer thickness and absorption coefficients are considered. Later, a SPICE circuit network is configured which in our case consist of 25 triple junction cells connected in series. The snippet of SPICE configured equivalent circuit where each cell consists of current sources, by-pass diodes and parasitic series and shunt resistances is shown in figure 6(b). The network is solved using SPICE circuit solver by applying a voltage from -10 V to 75 V across the network in steps of 0.1 V and calculating the corresponding current flowing through the network.

Figure 7 shows the experimental and simulated values of $V_{oc}$ and $I_{sc}$ for a modelled cell, before configuring the SPICE network. The value of measured $V_{oc}$ shown in figure 7 is the module $V_{oc}$ divided by number of cells, which in our case is 25. In the simulation, the input cell temperature to SolCore is the average of all channel cell temperatures. An offset value of temperature needs to be added to the measured average cell temperature as the cells are hotter than the temperature measured in the module behind the cell. The offset value of temperature was calculated by comparing the measured and simulated module $V_{oc}$ and empirically related to DNI (greater than 600 W/m²) and is of the form,

$$Offset\ temperarure\ (Celcius) = 0.029 \times DNI\left(\frac{W}{m^2}\right) - 17 \quad Eq.1$$

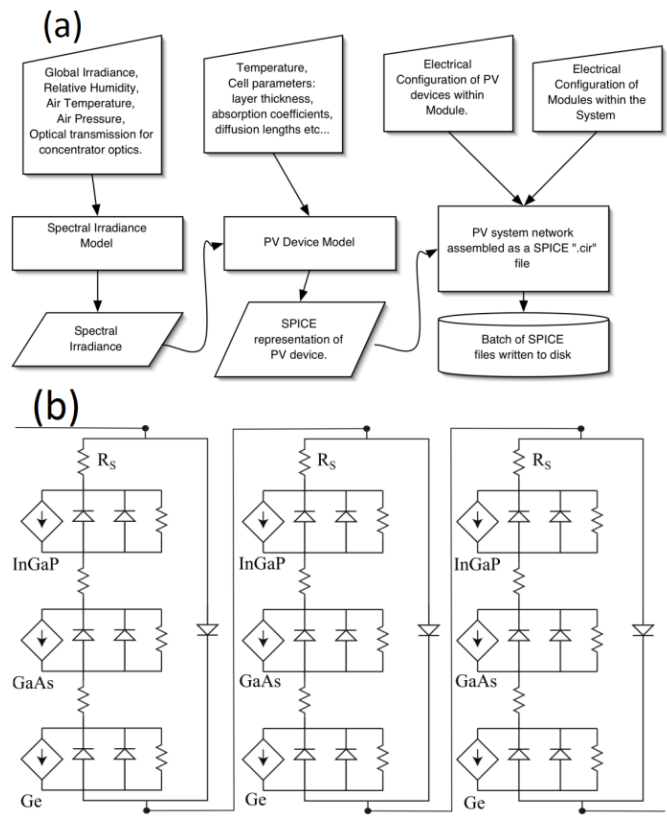

Fig. 6. (a) Bottom-Up approach for CPV simulation and (b) Schematic diagram snippet of the circuit network in SPICE

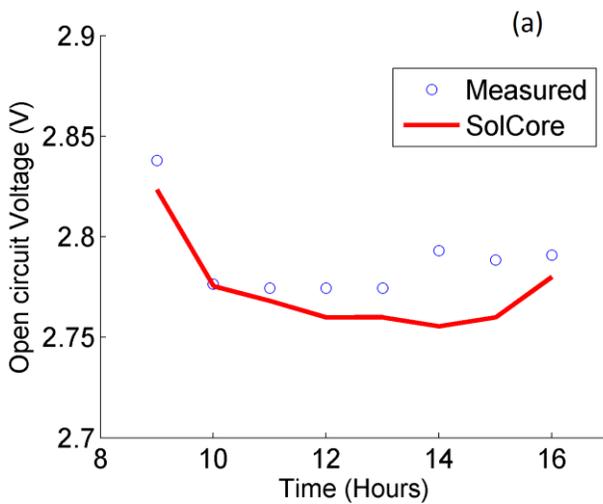

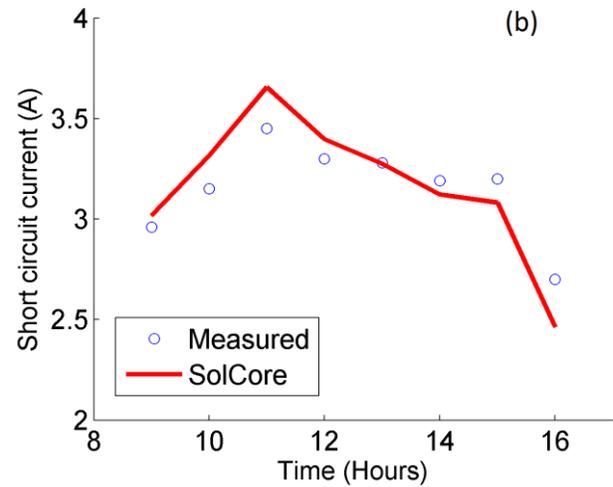

Fig. 7. Measured and Simulated values of (a) Open-Circuit voltage and (b) Short-circuit current on 7[th] January 2014

Because of current mismatches in module, the IV curve shows a gradual reduction in current with increasing bias and hence a poor fill-factor. To simulate the module with the cell current mismatch, our model uses a Gaussian distribution of short circuit currents. The value of variance for this distribution (which characterise mismatches of cell currents within the module) is fitted for the test module. The model is validated in Bangalore and figure 8 shows the measured and modelled IV curves.

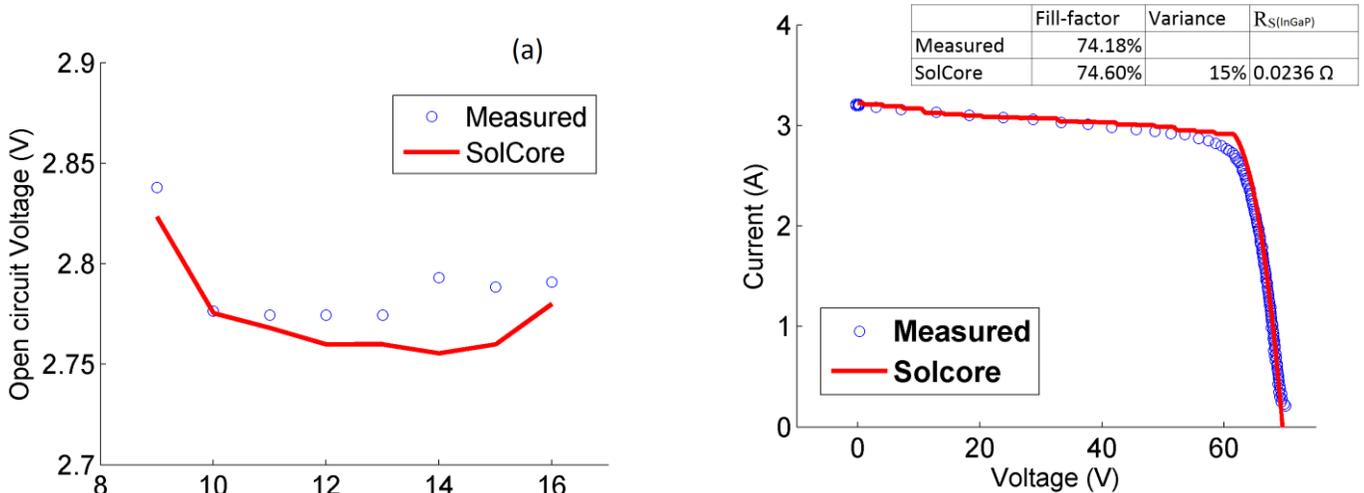

Fig. 8. Measured and modelled IV curves on 20[th] December 2012 at 10:55 AM

After decoupling the effects of spectrum on CPV performance, a fault was identified in the CPV module; here we consider anything that reduces the performance of CPV module beyond the expected performance limits under stable conditions (clear sky and clean optics) as a fault. The IV curve on 7[th] January 2014 at 1 PM is shown in figure 9 (After the fault occurred, the fill-factor reduced to about 70% while the fill-factor without the fault is 79%). By adjusting the distribution in $I_{sc}$, the cell current behaviour within the module could be modelled and compared with the IV curve expected with a much narrower distribution of $I_{sc}$ which is more typical of normal module behaviour; the frequency distributions of $I_{sc}$ used are shown in Figure 10.

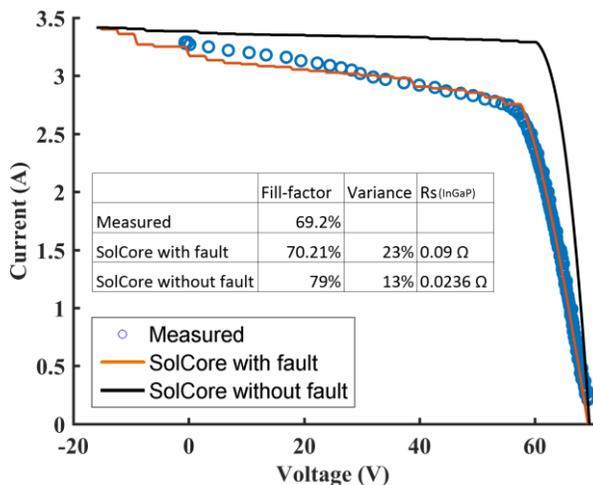

Fig. 9. Measured and modelled IV curves on 7$^{th}$ January 2014 at 1 PM (After the occurrence of fault)

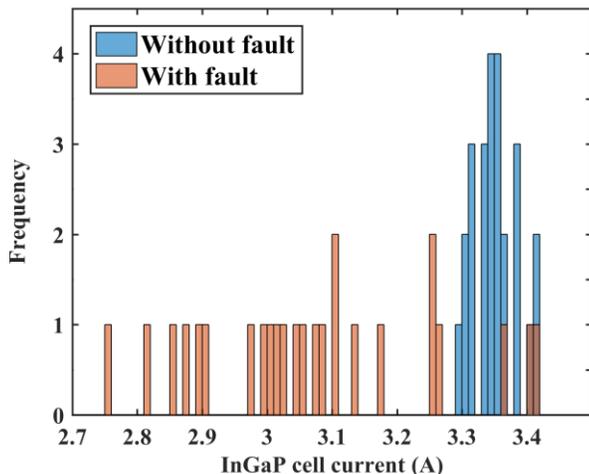

Fig. 10. Histogram of InGaP sub-cell currents in the module for IV curves shown in figure 9

### V. FAULT DIAGNOSIS

A fault in the module will reduce the energy output of the module. There might be many types of fault, for instance, increase in series resistance, reduction in shunt resistance, breakdown of by-pass diode, misalignment of module and high mismatch of cell currents within the module caused due to cell or optical degradation. Electrically resistive faults lead to a linear gradient in the IV curve close to $I_{sc}$ and $V_{oc}$ for respective shunt and series resistances. Our analysis fits the entire IV curve and can therefore provide additional information about the nature of the fault, beyond identifying whether it is purely resistive including mismatch losses in a string of CPV cells. For instance, in figure 9, the absolute value of the gradient $\frac{dI}{dV}$ close to $V_{oc}$ for simulated IV curves are 0.32 S and 0.72 S for with and without the fault respectively while the measured absolute gradient is 0.28 S. It should also be noted that high mismatch of cell currents within the module could be either due to soiling of the module causing non-uniform illuminations or tracker mechanical errors/misalignment of module. However, in our case, the module was cleaned every day and tracker accuracy was checked periodically. With the method proposed here, we can diagnose the module, even when there is no previous measured data about performance available for comparison.

After identifying the fault in our module, we modelled the fault and a comparison of power with and without fault is shown in figure 11. Here, the power without fault is simulated by turning off the fault in model. Further, simulations with the faulty and healthy conditions were run for all clear days of January, February and March 2014, when the spectral irradiance data was available and table 4 presents the daily average energy yield with and without fault together with the measured data.

The presence of faults can be identified by analysing the IV curves that are simulated using the equivalent circuit model without taking the module apart, in which case, we risk damage to the module. For example, a situation with large current mismatch in the module results in rapid degradation of current at lower bias conditions in IV curves. The flowchart of the algorithm for performance analysis and fault diagnosis is shown in Figure 12.

TABLE IV
MEASURED AND SIMULATED DAILY AVERAGE ENERGY YIELD

|  | Measured clear sky DNI (kWh/m$^2$/day) | Clear sky DNI - SMARTS2 modelled (kWh/m$^2$/day) | Daily average Measured yield (kWh/day) | Daily average modelled yield - SolCore with Fault (kWh/day) | Daily average modelled yield - SolCore without Fault (kWh/day) |
| --- | --- | --- | --- | --- | --- |
| January | 6.1 | 6.12 | 1.01 | 1.03 | 1.53 |
| February | 5.6 | 5.73 | 0.95 | 0.99 | 1.39 |
| March | 6.3 | 6.30 | 1.16 | 1.21 | 1.54 |

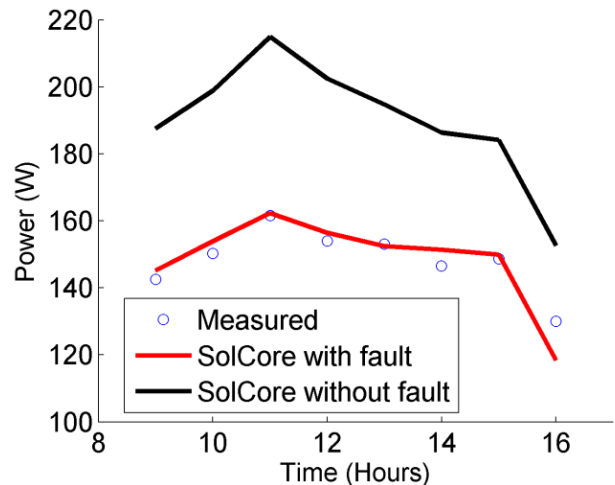

Fig. 11. Measured and modelled power on 7$^{th}$ January 2014

The algorithm can be triggered when the fill-factor of the concentrator module falls below certain lower threshold value, depending on module (In the present case, it is about 75%). The algorithm basically tries to fit the $I_{MP}$ (operating current point) by changing parameters that influence low bias current response and $V_{MP}$ (operating voltage point) by changing the parameters that influence the high bias current response. The algorithm tries to fit the $I_{MP}$ solely by increasing the mismatch in cell currents within the module (It can do so by increasing the variance of Gaussian curve, refer section 3). Increasing the mismatch would result in rapid degradation in current at lower bias in the IV curves. After fitting the IV curve at lower bias where the shunts and cell current mismatches are prominent, the algorithm looks at IV curve at higher bias where the series resistance is prominent. The algorithm checks if the value of $V_{MP}$ fits with 5% offset. If it does not, the algorithm increases the value of series resistance of all the current limiting cells (InGaP in our case) until it fits the $V_{MP}$. This increase in series resistance could have originated from interconnection of cells, from a crack in solder junction or thermal stress. However, the method discussed in this paper lumps these resistances and replaces it with an incremental resistance ΔR connected in series with the InGaP cell. Thus, increasing series resistance of InGaP cell need not necessarily mean that the actual cell resistance is increased but is a way to account for losses.

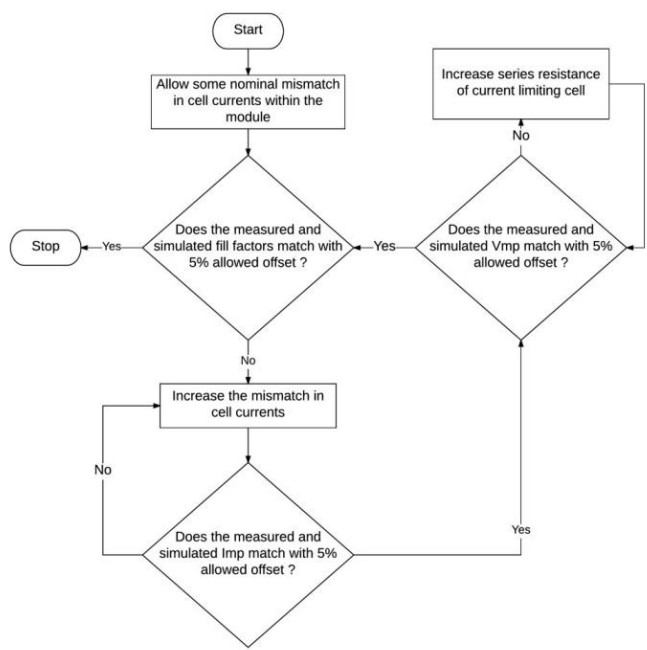

Fig. 12. Flowchart for performance assessment and fault diagnosis

In the module under test in Bangalore, by analysing the histogram of IV fitted InGaP sub-cell currents, we find that there is huge mismatch in cell currents within the module. Further, by looking at the current response at high electrical bias at different DNI values and comparing the corresponding measured IV curves, we learnt that the series resistance has increased. As a consequence of these effects, the performance of the module is reduced. By using the algorithm under different atmospheric conditions, we could model the mismatch in $I_{sc}$ by setting the value of variance of Gaussian curve at 0.23. Also, the algorithm tells us that the value of series resistance ($R_S$) of all InGaP cells must be increased by 0.0664 Ω. When we introduced the above-mentioned faults (variance and series resistance) in the model, we are getting the results that are consistent to measured quantities with less offset (See figure 11 and table 4).

## VI. CONCLUSION

The incident spectral irradiance on a CPV module in Bangalore is constructed from meteorological parameters using the radiative transfer code SMARTS2, resulting in generally good agreement with the measured spectrum. While the effects of spectral mismatch on short circuit current of a CPV module are apparent in the module power output, this effect is dwarfed by the presence of significant current mismatch in the module. These module faults have been characterized primarily by introducing a wide variation of short circuit current in the module resulting in an acceptable match between calculated and measured results. Finally, a method to diagnose faults in CPV modules is discussed.

ACKNOWLEDGEMENT

The authors gratefully acknowledge the supply of the CPV module from Daido Steel as part of the NGCPV project. This work has been partially supported by the European Commission and NEDO through the funding of the project NGCPV EUROPE-JAPAN (EU Ref. N: 283798).

REFERENCES

[1] N. L. A. Chan, H. E. Brindley, N. J. Ekins-Daukes, "Impact of individual atmospheric parameters on CPV system power energy yield and cost of energy", *Prog. Photovolt. Res. Appl.*, vol. 22, no. 10, pp. 1080-1095, 2014.

[2] M. Meusel, R. Adelhelm, F. Dimroth, A. W. Bett, and W. Warta, "Spectral mismatch correction and spectrometric characterization of monolithic III-V multi-junction solar cells," *Prog. Photovolt., Res. Appl.*, vol. 10, no. 4, pp. 243–255, 2002.

[3] K. Araki, M. Yamaguchi, "Influences of spectrum change to 3-junction concentrator cells", *Sol. Energy Mater. Sol. Cells*, vol. 75, pp. 707-714, 2003.

[4] Hashimoto J, Kurtz S, Sakurai K, Muller M, Otani K, "Field experience and performance of CPV system in different climates", *Proc. AIP Conf.*, vol. 1556, pp. 261–265, 2013.

[5] Young, T. B., A. Chan, N. J. Ekins-Daukes, K. Araki, Y. Kemmoku, and M. Yamaguchi. "Atmospheric considerations when estimating the energy yield from III-V photovoltaic solar concentrator systems." Presented at *25th European Photovoltaic Solar Energy Conference and Exhibition/5th World Conference on Photovoltaic Energy Conversion*, 2010.

[6] N. Chan, T. Young, H. Brindley, B. Chaudhuri, N. J. Ekins-Daukes, "Variation in spectral irradiance and the consequences for multi-junction concentrator photovoltaic systems", *Proc. 35th IEEE Photovoltaic Spec. Conf.*, pp. 003008-003012, 2010.

[7] Ekins-Daukes, N. J., Y. Kemmoku, K. Araki, T. R. Betts, R. Gottschalg, D. G. Infield, and M. Yamaguchi. "The design specification for Syracuse; a multi-junction concentrator system computer model", *Proc. 19th European Photovoltaic Solar Energy Conference*, pp. 2466-2469, 2004.

[8] N. Chan, T. B. Young, H. E. Brindley, N. Ekins-Daukes, K. Araki, Y. Y. Kemmoku, "Validation of energy prediction method for a concentrator photovoltaic module in Toyohashi Japan", *Prog. Photovoltaics Res. Appl.*, vol. 21, pp. 1598-1610, 2012.

[9] D. King, B. Hansen, J. Kratochvil, M. Quintana, "Dark current-voltage measurements on photovoltaic modules as a diagnostic or manufacturing tool", *Proc. 26th IEEE Photovoltaic Spec. Conf.*, pp. 1125-1128, 1997.

[10] Rummel S.R, McMahon T.J, "Effect of cell shunt resistance on module performance at reduced light levels", *In: Proceedings of the 13th NREL Photovoltaics Program Review*, pp. 581-586, 1995.

[11] M. P. P. Snchez Friera, J. C. J. Pelez, M. S. de Cardona, "Analysis of degradation mechanisms of crystalline silicon PV modules after 12 years of operation in Southern Europe", *Prog. Photovoltaics: Res. Appl.*, vol. 19, pp. 658-666, 2011.

[12] D. C. Jordan, S. R. Kurtz, K. T. VanSant, J. Newmiller, "Compendium of photovoltaic degradation rates", *Progress Photovolt. Res. Appl.*, vol. 24, no. 7, pp. 978-989, 2016.

[13] R. Platon, J. Martel, N. Woodruff, and T. Y. Chau, "Online fault detection in PV systems," *IEEE Trans. Sustain. Energy*, vol. 6, no. 4, pp. 1200–1207, Oct. 2015.

[14] E. van Dyk, E. Meyer, "Assessing the reliability and degradation of photovoltaic module performance parameters", *IEEE Trans. Reliabil.*, vol. 53, no. 1, pp. 83-92, Mar. 2004.

[15] Araki, Kenji, Hisafumi Uozumi, Toshio Egami, Masao Hiramatsu, Yoshinori Miyazaki, Yoshishige Kemmoku, Atsushi Akisawa, N. J. Ekins-Daukes, H. S. Lee, and Masafumi Yamaguchi. "Development of concentrator modules with dome-shaped Fresnel lenses and triple-junction concentrator cells.", *Prog. Photovoltaics Res. Appl.*, vol. 13, no. 6, pp. 513-527, 2005.

[16] Specification Sheet HCPV Modules: DACPV-280W25, Daido Steel

[17] Y. Ota, H. Nagai, K. Araki, K. Nishioka, "Temperature distribution in 820X CPV module during outdoor operation", *Proc. AIP Conf.*, vol. 1477, no. 1, pp. 364-367, 2012.

[18] Mitra, Indradip, Kaushal Chhatbar, Ashvini Kumar, Godugunur Giridhar, Ramdhan Vashistha, Richard Meyer, Marko Schwandt. "Solmap: Project in India's Solar Resource Assessment." *Int J Renew Energy Dev*, vol. 3, no. 3, pp. 207-216, 2014.

[19] Y. Ota, T. Sueto, H. Nagai, K. Araki, and K. Nishioka, "Reduction in Operating Temperature of 25 Series-Connected 820X Concentrator Photovoltaic Module," *Japanese Journal of Applied Physics*, vol. 52, no. 4S, 2013.

[20] M. Castro, C. Domínguez, R. Núñez, I. Antón, G. Sala, and K. Araki, "Detailed effects of wind on the field performance of a 50 kW CPV demonstration plant", *Proc. AIP Conf, vol.*1556, pp. 256-260, 2013

[21] C. Gueymard, "Parameterized transmittance model for direct beam and circumsolar spectral irradiance", *Solar Energy*, vol. 71, no. 5, pp. 325-346, 2001.

[22] R. Bird and C. Riordan, "Simple Solar Spectral Model for Direct and Diffuse Irradiance on Horizontal and Tilted Planes at the Earth's Surface for Cloudless Atmospheres", *Journal of Climate and Applied Meteorology*, vol. 25, no. 1, pp. 87-97, 1986.

[23] D. Alonso-Álvarez, T. Wilson, P. Pearce, M. Führer, D. Farrell and N. Ekins-Daukes, "Solcore: a multi-scale, Python-based library for modelling solar cells and semiconductor materials", *Journal of Computational Electronics*, vol. 17, no. 3, pp. 1099-1123, 2018.

[24] Ekins-Daukes, N.J., Betts, T.R., Kemmoku, Y., Araki, K., Lee, H.S., Gottschalg, R., Boreland, M.B., Infield, D.G. and Yamaguchi, M., 2005, January. "Syracuse-A multi junction concentrator system computer model", *Proc. 31st IEEE Photovoltaic Specialists Conference*, pp. 651-654, 2005.